%% file: transverse_loading_nuovo.tex
\documentclass[10pt,a4paper]{article}

\usepackage[T1]{fontenc}
\usepackage[utf8]{inputenc}
\usepackage{amssymb,amsmath,amsthm,amsfonts}    
\usepackage{bm}                                 
\usepackage{mathtools}                          
\usepackage{stmaryrd}                           
\usepackage{breqn}                              
\usepackage[textfont=footnotesize,labelfont={footnotesize,bf}]{caption}
\usepackage{subcaption}
\usepackage{booktabs}
\usepackage[inline]{enumitem}
\usepackage{graphicx}
\usepackage[dvipsnames]{xcolor}
\usepackage[top=30mm,right=30mm,left=30mm,bottom=30mm]{geometry}
\usepackage[colorlinks=true,linkcolor=blue,urlcolor=blue]{hyperref}
\usepackage[affil-it,auth-sc]{authblk}
\usepackage[colorinlistoftodos,textwidth=25mm,textsize=footnotesize]{todonotes}
\usepackage{lipsum}                             
\usepackage{circledsteps}
\usepackage[
	backend=biber,
	style=numeric,
	citestyle=numeric-comp,
	maxbibnames=99,
	minbibnames=99,
	maxcitenames=2,
	mincitenames=1,
	giveninits=true,
	sorting=none,
	sortlocale=auto,
	natbib=true,
	url=false,
	isbn=false,
	doi=true,
	eprint=true
]{biblatex}
\graphicspath{{./}{./figures/}}                 

\title{The strange mechanics of an elastic rod under null-resultant transverse loads}

\author[1]{Davide Bigoni\footnote{Corresponding author: e-mail: \href{mailto:bigoni@ing.unitn.it}{bigoni@ing.unitn.it}; phone: +39\,0461\,282507.}}
\author[2]{Diego Misseroni}
\author[1]{Andrea Piccolroaz}

\affil[1]{Instabilities Lab, University of Trento, Trento, Italy}
\affil[2]{Laboratory for the Design of Reconfigurable Metamaterials \& Structures,
DICAM, University of Trento, Italy}

\date{\today}

\begin{document}

\maketitle

\begin{abstract}
\noindent
Two equal and opposite distributed dead loads are applied orthogonally to the axis of an elastic rod in its rectilinear reference configuration, one at the extrados and the other at the intrados, such that the resultant applied force per unit length is uniformly zero. In this configuration, the rod is subjected to a transverse (tensile or compressive) stress, which is usually believed to have no significant effect on the structural response and has therefore not been considered so far. Contrary to this common belief, the asymptotic behavior of an incrementally deformed elastic layer and three different rod models (the first derived as an asymptotic approximation of the elastic layer; the second based on Euler elastica; and the third obtained by homogenization of a discrete model) reveal that this loading condition produces the same deformation in the rod as an axial load. 
In particular, the transverse load adds to the axial load in a generalized version of the Euler elastica, leading to buckling and nontrivial postcritical deformations when compressive. The critical transverse stress for buckling is found to have the same form as the Euler critical stress under axial force and tends to zero in the limit of vanishing rod inertia. For this reason, instability induced by transverse loading persists even when the rod thickness tends to zero. These theoretical predictions are confirmed by numerical simulations of a slender elastic layer, which show that increasing transverse load can induce buckling and drive the layer along a deformation path that closely follows that predicted by the generalized Euler elastica throughout the entire postcritical regime, even beyond self-intersection. 
To show that this behavior can be realized in practice, a dedicated experimental setup is developed, and the experimental results fully confirm the theoretical and numerical predictions.
The instability disclosed here may affect thin films and elastic layers subjected to transverse loading and is therefore relevant to several advanced technologies, including micro- and nanoscale devices. 
\end{abstract}

\paragraph{Keywords}
beam instability \textperiodcentered\
Euler buckling \textperiodcentered\
nonlinear elasticity

\section{Introduction}

\begin{figure}[!htb]
\centering
\includegraphics[width=0.9\textwidth]{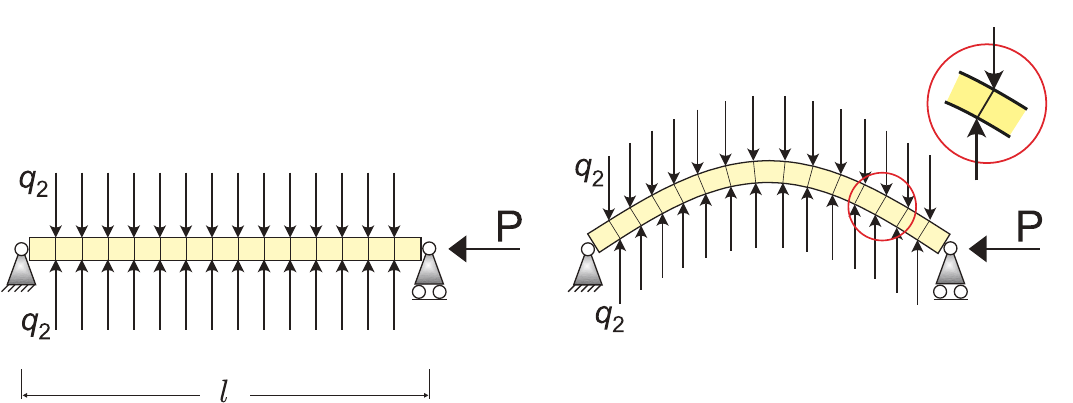}
\caption{
In addition to a compressive axial force $P$, a doubly supported elastic rod is subjected to two uniformly distributed loads $\pm q_2$ (illustrated with a negative sign) acting in opposite directions, one applied at the extrados and the other at the intrados of the rod. The loads $q_2$ are dead loads, meaning that their direction is fixed in the reference configuration (left) and their point of application moves with the deformation of the rod (right). The present article shows that the transverse loads $\pm q_2$ strongly influence the deformation of the rod, which is still governed by the Euler elastica, but with $q_2$ effectively added to the axial force $P$. Consequently, even the sole application of the transverse load may induce the structure to buckle and develop full postcritical behavior, as the deformation causes the transverse dead load to act as a couple distribution (see inset).
}
\label{appoggiata}
\end{figure}

A rod, straight and horizontal in its reference configuration, is subjected to two equal and opposite loads, simultaneously applied at its extrados and intrados, uniformly distributed along the axis, and acting orthogonally to it. With reference to Fig.~\ref{appoggiata} (where a compressive axial load $P$ also acts on a doubly supported rod), the two distributed loads $\pm q_2$ are assumed to be dead, so that they remain vertical and constant even when the rod undergoes nontrivial deformations, as shown in Fig.~\ref{appoggiata} on the right. 

Under these conditions, the rod is subjected to a uniform transverse stress with a null resultant, so that the load is self-equilibrated. The straight configuration is, therefore, a trivial equilibrium state, and the effects of such loading are commonly thought to leave the structure unaffected, as if it were unloaded.
This assumption follows from the classical Euler idealization, in which a rod, despite having bending stiffness, is treated as having zero thickness. As demonstrated in the present article, however, this idealization yields incorrect results when transverse loads with vanishing resultant are considered. In fact, it is shown that a sufficiently large transverse compressive load can induce buckling, followed by a postcritical response of the same type as that produced by a compressive axial force. Thus, both axial and transverse loads are destabilizing when compressive and stabilizing when tensile.
\footnote{
Note that this behavior is not related to the effect of the Poisson ratio, since at the onset of buckling the prestress state is uniaxial: either an axial prestress with $T_{11} \neq 0$ and $T_{22} = 0$, or a transverse prestress with $T_{11} = 0$ and $T_{22} \neq 0$.
}  

More precisely, it is shown that:  
\begin{quote}
{\it not only can a compressive transverse load induce buckling, but it also produces a non-trivial deflection identical to that caused by a compressive axial force, and therefore governed by the Euler elastica. In particular, the transverse stress adds to the axial stress in the following new form of the Euler elastica:}
\end{quote}
\begin{equation}
\label{ocio}
    \theta''-\frac{T_a+T_t}{E \rho^2}\sin{\theta} = 0,
\end{equation}
where $T_a = P/A$ and $T_t=q_2/b$ are the axial and transverse stresses, respectively, assumed positive when tensile, $E$ is the elastic modulus of the rod, $A$ its cross-section area, $b$ its out-of-plane thickness, and $\rho$ the radius of inertia of the cross-section. Note that the distributed load $q_2$ (and the axial load $P$) is defined as positive when it generates a tensile stress $T_t$ (respectively $T_a$), so that those reported in Fig.~\ref{appoggiata} are negative. Linearization of eq.~\eqref{ocio} yields the following bifurcation condition for a simply supported beam, providing a generalized buckling formula that includes the transverse stress:  
\begin{equation}
\label{ocio2}
    T_{a} + T_{t} = - n^2 \pi^2 \frac{E}{\lambda_{\text{sl}}^2}, \quad n= 1, 2, 3, ...
\end{equation}
where $\lambda_{\text{sl}} = l/\rho$ is the slenderness of the rod. Eqs.~\eqref{ocio} and \eqref{ocio2} show that axial stress $T_a$ and transverse stress $T_t$ {\it play exactly the same role}. 

Theoretically, the deformation and buckling of the transversely loaded rod, governed by eqs.~\eqref{ocio} and \eqref{ocio2}, are demonstrated in the present article through four different and independent approaches:

\begin{itemize}
    \item The bifurcation of an elastic (orthotropic and incompressible) layer in plane strain is analyzed following \cite{biot1965mechanics,  hill_bifurcation_1975, bigoni_effect_1997,  bigoni_nonlinear_2012}, and the limit as the thickness tends to zero is addressed. The layer is subjected to distributed dead loads of opposite signs at its upper and lower surfaces. 
    In this context, lateral expansion due to Poisson's effect does not play any role. 
    The critical transverse load for bifurcation is shown to converge to the equation \eqref{ocio2} in the limit of zero thickness-to-wavelength ratio.
    \item The governing equations for the same layer are shown, after linearization of the kinematics and use of mean stresses, to reduce to the same model describing the buckling of a straight axially prestressed beam, namely the linearization of the generalized Euler elastica, eq.~\eqref{ocio}.
    \item An Euler-Bernoulli rod is equipped with a finite thickness, and transverse dead loads are applied at its extrados and intrados. Upon rotation of the rod's axis, this loading generates a distribution of couples. The resulting structural model is shown to yield eq.~\eqref{ocio}.  
    \item A discrete chain of rigid links joined by elastic hinges is considered, following the approach initiated by Domokos \cite{domokos_qualitative_1993, velarde_odd_2002} and recently extended to complex microstructures  \cite{kocsis_discrete_2017,paradiso_nonlinear_2025}. Each link is equipped with a transverse rigid element that carries transverse dead loads at its ends. Homogenization of the model, as the link length tends to zero, yields eq.~\eqref {ocio}. 
\end{itemize}

The validity of eqs.~\eqref{ocio} and \eqref{ocio2} has two key consequences: (i.) {\it the nonlinear response of the elastica produced by transverse or axial loads is exactly the same}; (ii.) {\it in the limit of infinite slenderness, the buckling load tends to zero for both transverse and axial forces}. Therefore, the effect of transverse loading is fundamental and not a \lq spurious' artifact related to the rod thickness, in which case it would vanish as the thickness tends to zero. 

Although all four models above lead to identical conclusions, namely the validity of eqs.~\eqref{ocio} and \eqref{ocio2}, each relies on some form of idealization, either through asymptotic approximations or through the assignment of a finite thickness to the elastica. This raises the question of whether bifurcation and subsequent nonlinear postcritical behavior under transverse actions can actually occur in practice. To address this issue, two additional and independent validations are provided: numerical simulations and experimental investigations.\footnote{Videos explaining the main results of the article and the experiments are provided as supplementary material.}

The numerical simulations confirm the theoretical predictions and demonstrate that a transverse dead load, such as that illustrated in Fig.~\ref{appoggiata}, can drive a slender elastic layer along the deformation path predicted by the generalized Euler elastica, eq.~\eqref{ocio}, even beyond self-intersection. 

The experiments required the design and realization of a new experimental setup, in which an elastic rod is subjected to transverse tensile dead loads and tested under increasing axial compression. The experimental results not only validate the theoretical findings in an indisputable manner but also prove that the postulated loading condition can be effectively realized in practice.  

These results introduce a new paradigm for the bifurcation and postcritical behavior of rods and elastic layers, with potential implications for deformable manipulators subjected to transverse forces and for the mechanics of thin films, widely employed in micro- and nanotechnologies.

\section{A premise: bifurcation of an incompressible elastic layer under axial stress and transverse dead load}
\label{biotbifurc}

The incremental bifurcation of an incompressible elastic layer deformed in plane strain is a well-known and extensively investigated problem \cite{biot1965mechanics,hill_bifurcation_1975,bigoni_nonlinear_2012}. The layer is composed of an orthotropic elastic material and is subjected to a stress state defined by the Cauchy principal stresses $T_{11}$ and $T_{22}$. Its incremental behavior can be expressed through the following dimensionless parameters:
\begin{equation}
\label{nina}
    \xi = \frac{\mu^{*}}{\mu}, \quad 
    \eta = \frac{T_{11} + T_{22}}{2 \mu}, \quad 
    k = \frac{T_{11} - T_{22}}{2 \mu},
\end{equation}
where $\mu$ and $\mu^*$ are incremental shear moduli. The incremental nominal stress $\dot{t}_{ij}$ is related to the incremental displacement gradient $v_{i,j}$ and the incremental in-plane mean stress $\dot{p}$ by
\begin{equation}
\label{costicazzo}
    \begin{aligned}
        \dot{t}_{11} &= \mu (2\xi - k - \eta) v_{1,1} + \dot{p}, & 
        \dot{t}_{22} &= \mu (2\xi +k-\eta) v_{2,2} + \dot{p}, \\[3mm]
        \dot{t}_{21} &= \mu [(1+k) v_{2,1} + (1-\eta) v_{1,2}], &
        \dot{t}_{12} &= \mu [(1-\eta) v_{2,1} + (1-k) v_{1,2}], 
    \end{aligned}
\end{equation}
subject to the incompressibility condition $v_{1,1} + v_{2,2} = 0$. 

\begin{figure}[!htb]
\centering
\includegraphics[width=8cm]{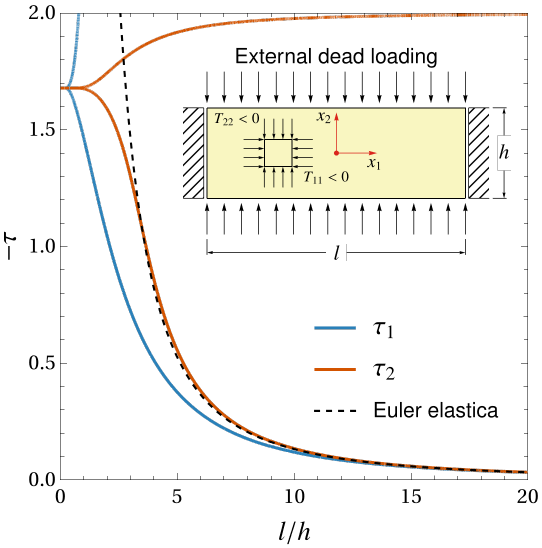}
\caption{
An elastic layer of current thickness $h$ and length $l$ (see inset), made of incompressible neo-Hookean material with incremental modulus $\mu$, is subjected to either a longitudinal or transverse {\it dead} loading, inducing a uniaxial prestress: longitudinal ($\tau_1<0$ with $\tau_2=0$) or transverse ($\tau_2<0$ with $\tau_1=0$). The bifurcation stresses $\tau_1$ and $\tau_2$, obtained from the solution of eqs.~\eqref{una} and \eqref{dua}, respectively, are plotted as functions of the layer slenderness $l/h$ (for the simply supported case $\zeta = \pi h/l$). The Euler buckling load is also shown as a dashed curve. Although the curves for $\tau_1$ and $\tau_2$ differ for thick layers, they converge at large slenderness, say $l/h$>5, both asymptotically approaching the Euler buckling load. Therefore, a transverse dead load has the same destabilizing effect as a longitudinal dead load on a slender elastic rod.
}
\label{fig01}
\end{figure}

Equilibrium requires 
\begin{equation}
\label{equili}
    \dot{t}_{11,1} + \dot{t}_{21,2} = 0, \quad 
    \dot{t}_{12,1} + \dot{t}_{22,2} = 0,
\end{equation}
together with boundary conditions. For a rectangular block centered in a Cartesian reference system, see the inset in Fig.~\ref{fig01}, the boundary conditions are:
\begin{itemize}
    \item For a {\it dead} loading transverse to the layer, vanishing incremental nominal tractions at both upper and lower faces:
    \begin{equation}
    \label{onebiot}
        \dot{t}_{22} \left( x_{1},x_2=\pm h/2 \right) = 0, \quad 
        \text{and} \quad 
        \dot{t}_{21} \left( x_{1},x_2=\pm h/2 \right)=0;
    \end{equation}
    \item For a longitudinal prestress, obtained through compression of the layer between rigid frictionless lateral constraints, vanishing incremental nominal shear reactions, and vanishing incremental horizontal displacement:
    \begin{equation}
    \label{secondbiot}
        \dot{t}_{12} \left( x_{1}=\pm l/2,x_2 \right) = 0, \quad
        \text{and} \quad
        v_{1} \left( x_{1}=\pm l/2,x_2 \right)=0.
    \end{equation}
\end{itemize}

The incremental displacement field is represented as \cite{bigoni_nonlinear_2012}
\begin{equation}
\label{zorro1}
    \begin{aligned}
    v_1 &= -\Big[ b_1 \Omega_1 e^{i c_1 \Omega_1 x_2} + b_2 \Omega_2 e^{i c_1 \Omega_2 x_2} + b_3 \Omega_3 e^{i c_1 \Omega_3 x_2} + b_4 \Omega_4 e^{i c_1 \Omega_4 x_2} \Big] e^{i c_1 x_1}, \\[3mm]
    v_2 &= \Big[ b_1 e^{i c_1 \Omega_1 x_2} + b_2 e^{i c_1 \Omega_2 x_2} + b_3 e^{i c_1 \Omega_3 x_2} + b_4 e^{i c_1 \Omega_4 x_2} \Big] e^{i c_1 x_1},
    \end{aligned}
\end{equation}
where $c_1$ is the wavenumber of the bifurcation mode, related to the wavelength $\lambda$ by 
\begin{equation}
    c_1 = 2 \pi / \lambda.
\end{equation}
Moreover, the in-plane mean stress increment is expressed as \cite{bigoni_nonlinear_2012}
\begin{multline}
\label{zorro2}
    \dot{p} = - i\mu c_1 \Big[ (k+\Lambda) (\Omega_1 b_1 e^{i c_1 \Omega_1 x_2} + \Omega_3 b_3 e^{i c_1 \Omega_3 x_2}) \\[3mm] 
    + (k-\Lambda) (\Omega_2 b_2 e^{i c_1 \Omega_2 x_2} + \Omega_4 b_4 e^{i c_1 \Omega_4 x_2}) \Big] e^{i c_1 x_1}. 
\end{multline}
The coefficients $b_j$ ($j=1,2,3,4$) in eqs.~\eqref{zorro1} and \eqref{zorro2} are constants, for the moment arbitrary, while the parameters 
\begin{equation}
    \Lambda = \sqrt{4\xi^2-4\xi+k^2}, \quad 
    \Omega_j = \pm\sqrt{\frac{1-2\xi+(-1)^j\Lambda}{1-k}}, 
\end{equation}
are functions of the prestress. 

A neo-Hookean material is assumed, for which
\begin{equation}
    \xi = 1, \quad 
    \Lambda = |k|, \quad 
    \Omega_1 = -\Omega_3 = i \sqrt{\frac{1+|k|}{1-k}}, \quad  
    \Omega_2 = -\Omega_4 = i \sqrt{\frac{1-|k|}{1-k}}. 
\end{equation}
The representations \eqref{zorro1} and \eqref{zorro2} automatically satisfy the equilibrium equations \eqref{equili}, so that their substitution into the constitutive equations \eqref{costicazzo} and a final imposition of the boundary conditions \eqref{onebiot} — the conditions \eqref{secondbiot} are automatically satisfied — yield a linear eigenvalue problem, which admits non-trivial solutions when the following two equations (which may appear to differ only by an exchange of indices, but in fact do not) are satisfied for axial and transverse prestress, respectively:
\begin{itemize}
    \item Axial prestress: $-2 < \tau_{1} < 2$, $\tau_{2} = 0$
    \begin{equation}
    \label{una}
        1 - \cosh \zeta \cosh \left( \zeta \sqrt{\frac{2 + \tau_1}{2 - \tau_1}} \right)
        + \frac{32 - 16\tau_1 + 4\tau_1^3 - \tau_1^4}{8(2-\tau_1) \sqrt{4-\tau_1^2}}
        \sinh \zeta \sinh \left( \zeta \sqrt{\frac{2 + \tau_1}{2 - \tau_1}} \right) = 0;
    \end{equation}
    \item Transverse prestress:$-2 < \tau_{2} < 2$, $\tau_1=0$
    \begin{equation}
    \label{dua}
        1 - \cosh \zeta \cosh \left( \zeta \sqrt{\frac{2 - \tau_2}{2 + \tau_2}} \right)
        + \frac{32 - 16\tau_2 + 4\tau_2^3 - \tau_2^4}{8(2-\tau_2) \sqrt{4-\tau_2^2}}
        \sinh \zeta \sinh \left( \zeta \sqrt{\frac{2 - \tau_2}{2 + \tau_2}} \right) = 0;
    \end{equation}
\end{itemize}
where $\zeta = c_1 h$ and $\tau_1 = T_{11}/\mu$, and $\tau_2 = T_{22}/\mu$ are the two dimensionless prestresses (positive in tension). 

Considering the elastic layer shown in the inset of Fig.~\ref{fig01}, where longitudinal displacements are constrained on the lateral boundaries by two frictionless rigid planes, the bifurcation wavelength $\lambda$ can be selected to model an elastic rod of length $l$ with different boundary conditions: (i.) $\lambda = l$, doubly clamped rod; (ii.) $\lambda = 2l$, simply supported rod (pinned-pinned or slider-slider); (iii.) $\lambda = 4l$, cantilever (clamped-free) or mixed pin-slider rod.

With reference to the doubly supported (or constrained with two opposite sliders as in the inset of the figure) scheme, $\lambda = 2l$, the bifurcation stresses $\tau_1$ and $\tau_2$ obtained from the solution of eqs.~\eqref{una} and \eqref{dua} are plotted in Fig.~\ref{fig01} against the slenderness ratio $l/h$. The figure also includes the Euler buckling load (dashed curve) for the corresponding simply supported beam. For transverse load $\tau_2$, bifurcation of a layer subjected to uniaxial transverse stress $T_{22}<0$ with $T_{11}=0$ is reported, while for longitudinal load $\tau_1$, the layer is subjected to uniaxial longitudinal stress $T_{11}<0$ with $T_{22}=0$.

For both axial and transverse loading, the solutions display two branches that emerge from $\tau_1 = \tau_2 \approx 1.67857$ approached in the limit $l/h \to 0$, corresponding to the onset of surface instability. Along the upper branches, the curves for axial and transverse loading remain distinct, so that the two types of prestress produce dissimilar effects. Along the lower branches, however, the situation changes: although the two curves differ significantly for thick layers ($l/h < 5$), they converge and eventually superimpose at high slenderness, recovering the Euler buckling load. Note that the transverse stress $\tau_2$ approaches the Euler curve more rapidly, remaining close to it over a broader range of slenderness values than the axial stress $\tau_1$. For sufficiently slender layers ($l/h>10$), both $\tau_1$ and $\tau_2$ coincide with the Euler bifurcation.

The fact that both the transversely and axially loaded layers tend to the Euler buckling load can be made explicit by expanding eqs.~\eqref{una} and \eqref{dua} in a Taylor series about $\zeta=0$. Truncated at fourth order, the expansion yields for eq.~\eqref{una}
\begin{equation}
\label{gattapaola1}
    \frac{\tau_1^2\, \zeta^2 \left[ 3\tau_1(4 - \tau_1)(2 - \tau_1) + 2\zeta^2\left(4 + \tau_1(2 - \tau_1)\right) \right]}{24(2 - \tau_1)^3} = 0,
\end{equation}
while for eq.~\eqref{dua}
\begin{equation}
\label{gattapaola2}
    \frac{\tau_2^2\, \zeta^2 \left[ 3\tau_2(4 - \tau_2)(2 + \tau_2) + 2\zeta^2\left(4 + \tau_2(2 - \tau_2)\right) \right]}{24(2 - \tau_2)(2 + \tau_2)^2} = 0. 
\end{equation}
Despite differing only by an exchange of indices, the two equations exhibit subtle sign differences.  
Finally, another series expansion of equations (\ref{gattapaola1}) and (\ref{gattapaola2}) about $\tau_1=0$ and $\tau_2=0$ leads to the critical loads for a wavelength $\lambda = 2l$ ($\zeta = \pi h/l$)
\begin{equation}
    \tau_1 = \tau_2 = -\frac{\zeta^2}{3} = -\frac{\pi^2 h^2}{3l^2}, 
\end{equation}
which coincides with the Euler buckling stress of a simply supported rod, $-\pi^2\bar{E}J/(l^2h)$, where $\bar{E}=4 \mu$ is the plane strain elastic modulus and $J=h^3/12$ the second moment of inertia. For other boundary conditions, the same asymptotic reasoning applies, yielding the corresponding Euler loads, e.g. $-4\pi^2\bar{E}J/(l^2h)$ for a clamped–clamped rod with $\lambda=l$, or $-\pi^2\bar{E}J/(4l^2h)$ for a cantilever with $\lambda=4l$. 

In conclusion, the bifurcation of an elastic layer under transverse dead load leads, in the slender limit, to the same Euler buckling stress as under axial dead load. This leads to the proposition that transverse dead loading of an elastic rod has the same destabilizing effect as axial loading, a conjecture that is rigorously confirmed in the next section.

\section{Three different approaches leading to the elastica and buckling for a transversely loaded rod}
\label{AnalyticalA}

The deformation and bifurcation of an elastic rod under a transverse dead load can be analyzed through three complementary approaches. Specifically, the reference theories are: (i.) the incremental asymptotic analysis of the deformation of an elastic layer similar to that considered in the previous section, (ii.) a modified Euler elastica model, where an Euler-Bernoulli rod is enhanced with a transverse thickness, and (iii.) a homogenized model derived from a discrete chain of rigid elements.

\subsection{Incremental asymptotics for bifurcation of an elastic layer subject to transverse dead load}
\label{kinekirchhoff} 

Consider an incompressible elastic layer of current thickness $h$ in the $x_1$-$x_2$ plane, where $x_1$ is the axial direction and $x_2 \in [-h/2,h/2]$. The layer is simultaneously subjected to an axial Cauchy stress $T_{11}$ and a transverse dead load $q_{2}$, which induces a transverse Cauchy stress $T_{22}$, Fig.~\ref{fig01}. 

Any incremental perturbation (denoted with a superimposed dot) satisfies the equilibrium equations \eqref{equili}. Through-thickness integration of equation \eqref{equili}$_1$, multiplied by $x_2$, gives
\begin{equation}
\label{socm}
    M_{,1} + m - \int_{-h/2}^{h/2}\dot {t}_{21}\, dx_{2} = 0,
\end{equation}
where the bending moment $M$ and the externally applied couples $m$ are defined as 
\begin{equation}
    M = \int_{-h/2}^{h/2}\dot {t}_{11} x_{2}\, dx_{2}, \quad  m =\dot {t}_{21} x_{2}  \Big|_{-h/2}^{h/2}. 
\end{equation}
Using the incremental constitutive equations \eqref{costicazzo}
\begin{equation}
    \dot {t}_{21} - \dot {t}_{12} = - T_{11} v_{2,1} + T_{22} v_{1,2} ,
\end{equation}
and introducing the shear force 
\begin{equation}
    S = \int_{-h/2}^{h/2}\dot {t}_{12}\, dx_{2}  ,
\end{equation}
equation \eqref{socm} becomes 
\begin{equation}
\label{integra1}
    M_{,1} + m - S + T_{11} h \overline{v}_{2,1} - T_{22} \Big [ v_{1} \left( h/2 \right) - v_{1} \left(- h/2 \right) \Big ] = 0
\end{equation}
in which the mean transverse displacement has been introduced
\begin{equation}
    \overline{v}_{2} = \frac{1}{h} \int_{-h/2}^{h/2} v_{2}\, dx_{2}.
\end{equation}

Integration of equation \eqref{equili}$_2$ through the thickness yields 
\begin{equation}
    S_{,1} + \dot {t}_{22}  \Big|_{-h/2}^{h/2}= 0,
\end{equation}
so that, under dead loading, $\dot {t}_{22} \left( \pm h/2 \right)=0$, it follows that $S_{,1}=0$. Hence, equation \eqref{integra1} reduces to
\begin{equation}
    M_{,11}+ T_{11} h \overline{v}_{2,11} - T_{22} \left[ v_{1,1} \left(x_1, h/2 \right) - v_{1,1} \left(x_1,- h/2 \right) \right] = 0, 
\end{equation}
{\it an expression that is exact}.

Adopting the standard linearized kinematic assumption of rod theory
\begin{equation}
    \overline{v}_{2} = w \left( x_{1} \right) , \quad v_{1}=u_{0} \left( x_{1} \right) - x_{2} w_{,1},
\end{equation}
one finds 
\begin{equation}
    \overline{v}_{2,1}+v_{1,2}=0.    
\end{equation}
For an incompressible material, $v_{2,2}+v_{1,1}=0$, the constitutive response reduces to
\begin{equation}
\label{costitazzo}
    \dot{t}_{11} = \left[ 4 \mu_{*} - \left( T_{11} + T_{22} \right) \right] v_{1,1} .
\end{equation}

From eq.~\eqref{costitazzo}, the bending moment $M$ then becomes 
\begin{equation}
    M = - D\, w_{,11}
\end{equation}
where $D$ represents the bending stiffness of the rod, a function of the second moment of inertia of the cross section, $h^3/12$, the incremental shear modulus, $\mu_*$, and the stress state as 
\begin{equation}
\label{DDD}
    D =  \frac{h^{3}}{12} \left[ 4 \mu_{*} - \left( T_{11} + T_{22} \right) \right] .
\end{equation} 
Finally, noting that 
\begin{equation}
    v_{1,1} \left(x_1,h/2 \right) - v_{1,1} \left(x_1,- h/2 \right) = - h\, w_{,11},
\end{equation}
{\it the differential equation governing the asymptotic quasi-static deformation of a straight layer is obtained as} 
\begin{equation}
\label{figata}
    w_{,1111} - h \frac{ T_{11} + T_{22} }{D} w_{,11}= 0. 
\end{equation}
Equation \eqref{figata} is formally identical to the linearized version of the elastica \eqref{ocio}$_1$, valid for a linear elastic beam prestressed with an effective force $h(T_{11}+T_{22})$. In the classical theory, the axial load is $hT_{11}$, but here it is evident that the transverse stress $T_{22}$ plays the same role as $T_{11}$, except that $hT_{22}$ is not the resultant of the transverse load, which is null. Accordingly, equation \eqref{figata} leads to the Euler buckling condition \eqref{ocio}$_2$. Remarkably, {\it the two stress components, $T_{11}$ and $T_{22}$, contribute in the same way to buckling}.

\subsection{The Euler elastica with thickness: instability under transverse dead loading}
\label{ELASTICAtheory}

\begin{figure}[!htb]
\centering
\includegraphics[width=8cm]{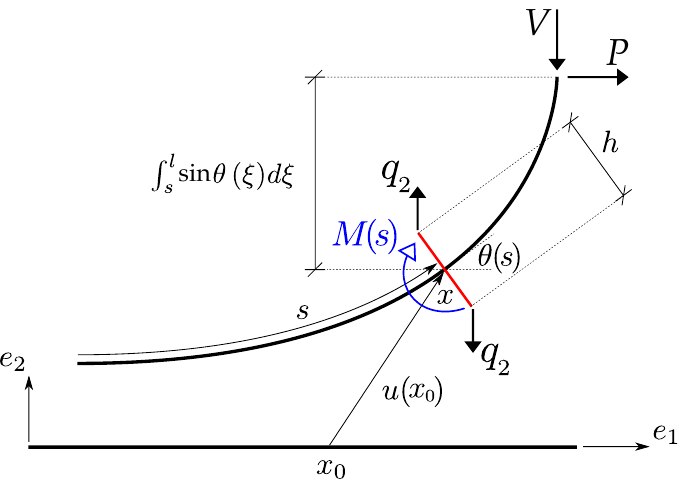}
\caption{
A model of elastica with a transverse cross-section of height $h$, which defines the extrados and intrados, where the uniform dead loads $q_2$ (shown positive) are applied.
}
\label{elasticaza}
\end{figure}

The governing equation of the Euler elastica can be re-derived by explicitly accounting for the rod's finite thickness. The thickness defines the extrados and intrados, at which a uniform transverse dead load $q_2$ is applied, Fig.~\ref{elasticaza}. Because the load $q_2$ is dead, it provides a distributed bending moment whenever the rod is curved. In addition to $q_2$, two end loads, an axial force $P$ and a shear force $V$, are considered. 

The external bending moment at the arclength coordinate $s$ is
\begin{equation}
\label{Mext}
    \mathcal{M}(s) = -\left( P  + q_{2} h \right) \int_{s}^{l} \sin \theta(\xi)\, d\xi - V \int_{s}^{l} \cos \theta(\xi)\, d\xi,
\end{equation}
which must equal the internal bending moment generated by the curvature, $EJ\, \theta'(s)$, where $EJ$ is the bending stiffness of the rod. This gives
\begin{equation}
\label{MINTEXT}
    \theta'(s) + \frac{P  + q_{2} h}{EJ} \int_{s}^{l} \sin \theta(\xi)\, d\xi + \frac{V}{EJ} \int_{s}^{l} \cos \theta(\xi)\, d\xi = 0.  
\end{equation}
Differentiating with respect to $s$ yields a new form for the Euler elastica 
\begin{equation}
\label{ELASTICA}
    \theta''(s) - \frac{P + q_{2} h}{EJ} \sin{\theta(s)} - \frac{V}{EJ} \cos{\theta(s)} = 0.
\end{equation}  
Eq.~\eqref{ELASTICA} represents the Euler elastica generalized to include the transverse dead loads per unit length $q_2$, in addition to the axial force $P$. The distributed transverse loading $q_2$ enters in the same way as the axial load $P$, effectively reducing (or increasing, if tensile) the critical force for instability.

\subsection{A homogenization approach to the elastica under transverse load}
\label{homo}

A discrete structure mimicking the elastica is analyzed under transverse dead loading, and homogenization is finally performed in the limit as the chain elements shrink to vanishing length. In this way, the continuum elastica equation \eqref{ELASTICA} is recovered. The approach follows the spirit of Domokos \cite{domokos_qualitative_1993, velarde_odd_2002} and subsequent developments \cite{kocsis_discrete_2017, paradiso_nonlinear_2025}, which use discrete micromechanical models to capture complex beam-like behaviors. 

\begin{figure}[!htb]
\centering
\includegraphics[width=.7\textwidth]{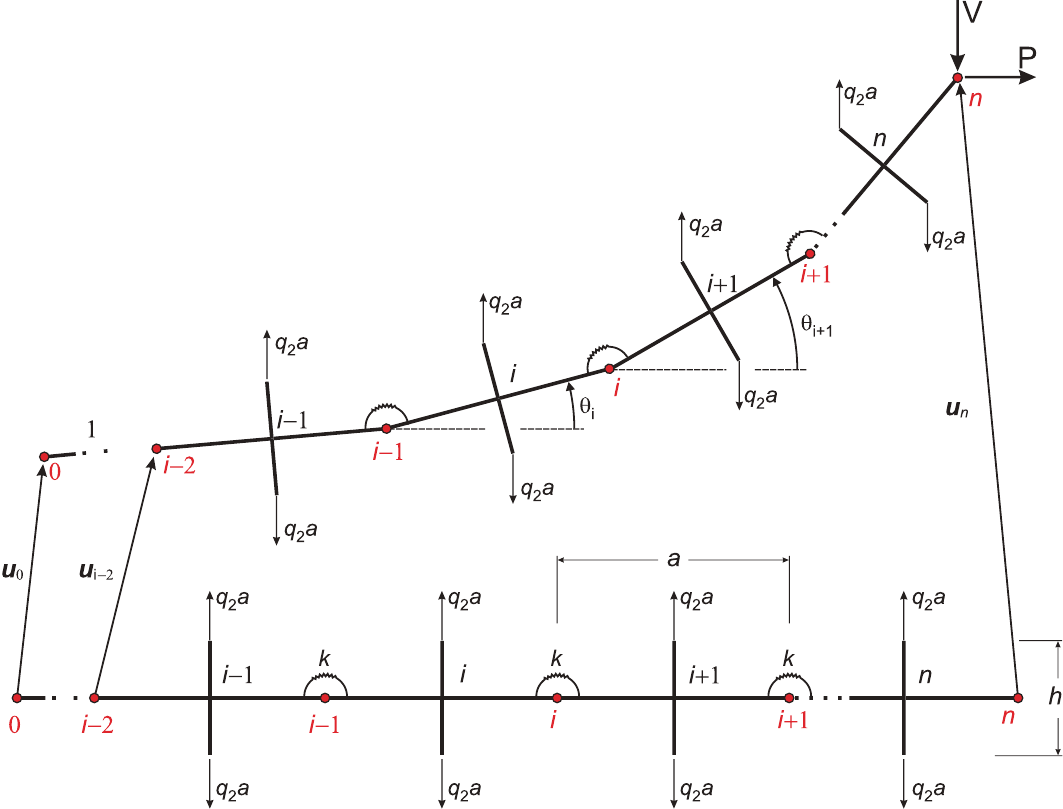}
\caption{
Micromechanical discrete model of $n$ rigid elements (of length $a$) connected by elastic hinges of stiffness $k$, mimicking an elastic rod of thickness $h$. There are $n+1$ nodes (numbered from $0$ to $n$), with the last node on the right subjected to two end forces $P$ (horizontal) and $V$ (vertical), both shown positive. Each element carries equal and opposite transverse dead loads $Q = q_2 a$ (shown positive).
}
\label{struttura_homogenizzazione}
\end{figure}

A chain of rigid elements of length $a$ is considered, connected by rotational springs of stiffness $k$. Differently from Domokos' model \cite{domokos_qualitative_1993}, each element is also equipped with a rigid transverse bar of height $h$, allowing the transmission of a transverse dead load and thus introducing thickness into the model, Fig.~\ref{struttura_homogenizzazione}. The chain, consisting of $n$ rigid elements, has a total length of $l = n\, a$.

Dead loads $P$ (horizontal, positive when tensile) and $V$ (vertical, positive when downward) are applied at the right end of the chain, while each element is transversely loaded by equal and opposite dead forces $Q = q_2 a$. When the $i$--th element rotates by an angle $\theta_i$, the total potential energy $\mathcal{P}$ of the chain is 
\begin{equation}
    \mathcal{P} = \frac{k}{2} \sum_{i=1}^{n-1} (\theta_{i+1} - \theta_{i})^2 + 
    a (P + q_2 h) \Big( n - \sum_{i=1}^n \cos \theta_i \Big) + a V \sum_{i=1}^n \sin \theta_i.
\end{equation}
The stationarity of $\mathcal{P}$ yields the equilibrium conditions
\begin{equation}
    \frac{\partial \mathcal{P}}{\partial \theta_i} = 
    k (2\theta_{i} - \theta_{i+1} - \theta_{i-1}) + a \Big[ (P+q_2h) \sin \theta_i + V \cos \theta_i \Big] = 0, \quad 
    i = 1, \cdots, n,
\end{equation}
or equivalently
\begin{equation}
\label{wess}
    \frac{\theta_{i+1} - 2\theta_{i} + \theta_{i-1}}{a^2} - \frac{P+q_2h}{k a} \sin \theta_i - \frac{V}{k a} \cos \theta_i = 0, \quad 
    i = 1, \cdots, n,
\end{equation}
In the limit $a \to 0$, the product $ka$ is kept finite by letting $k \to \infty$. The system thus homogenizes to the continuous equation
\begin{equation}
\label{wess2}
    \theta'' - \frac{P+q_2h}{k a} \sin \theta - \frac{V}{k a} \cos \theta = 0,
\end{equation}
where $\theta=\theta(s)$ is now a continuous function of the arclength $s$, describing a deformed elastica. 

Comparison of eq.~\eqref{wess2} with the elastica equation \eqref{ELASTICA} shows that the discrete chain homogenizes into a continuous rod with effective bending stiffness $ka$. Clearly, the transverse load $q_2h$ enters the governing equation in the same way as the axial force $P$, thus leading again to the Euler buckling condition \eqref{ocio}$_2$. 


\section{Post-buckling of a slender elastic layer: numerical vs elastica}
\label{post}

A numerical experiment using finite element simulations is presented in this section to validate the theoretical developments introduced above. In particular, the verification of eq.~\eqref{ocio} requires that self-equilibrated transverse loads deform the rod in the same way as an axial load. To this end, the perfect Euler elastica subjected to an axial stress $T_a= \sigma$ is compared to a slender elastic layer that is loaded with a transverse stress of the same magnitude $T_t=\sigma$.

The finite element simulations were carried out using the software Comsol Multiphysics. The structure is modeled as a two-dimensional plane-strain body with a prismatic beam geometry, a length of $l=30$ m, and a square cross-section ($b=h=1$ m). The material is assumed to be linear elastic, with Young’s modulus $E=210$ GPa and Poisson’s ratio $\nu=0.3$. The computational domain is discretized using quadratic quadrilateral serendipity elements in a structured mesh, resulting in $900$ elements ($150$ along the longitudinal direction and $6$ in the transverse direction, with an aspect ratio height/width of $0.833333$) and $6026$ degrees of freedom.
\footnote{
A finer mesh comprising $4320$ elements ($360$ along the longitudinal direction and $12$ in the transverse direction, with unit aspect ratio) and $27410$ degrees of freedom yields identical results.
}

Geometric nonlinearity is included to capture large-deflection effects relevant to buckling and post-buckling behavior. To activate buckling in the nonlinear analysis, the geometry is perturbed with an initial imperfection corresponding to the first buckling mode, with maximum midspan deflection scaled to $0.225$ m. Without this perturbation, the perfect structure follows the fundamental equilibrium path and does not exhibit bifurcation. Boundary conditions reproduce a simply supported beam, so that the mid-axis points at the two end sections are constrained in the transverse direction, while the mid-axis point at midspan is constrained in the axial direction to prevent rigid-body motion. The loading consists of self-equilibrated distributed line forces applied along the upper and lower edges of the cross-section, producing a uniform transverse compressive stress component $T_{t}$.

Two simulation studies are performed:
\begin{itemize}
    \item A linear buckling analysis is conducted to determine the critical buckling load and its associated eigenmode.
    \item A geometrically nonlinear analysis, incorporating the imposed initial imperfection, is performed to trace the post-buckling response under increasing transverse compressive forces.
\end{itemize}

\begin{figure}[!htb]
\centering
\includegraphics[width=0.9\linewidth]{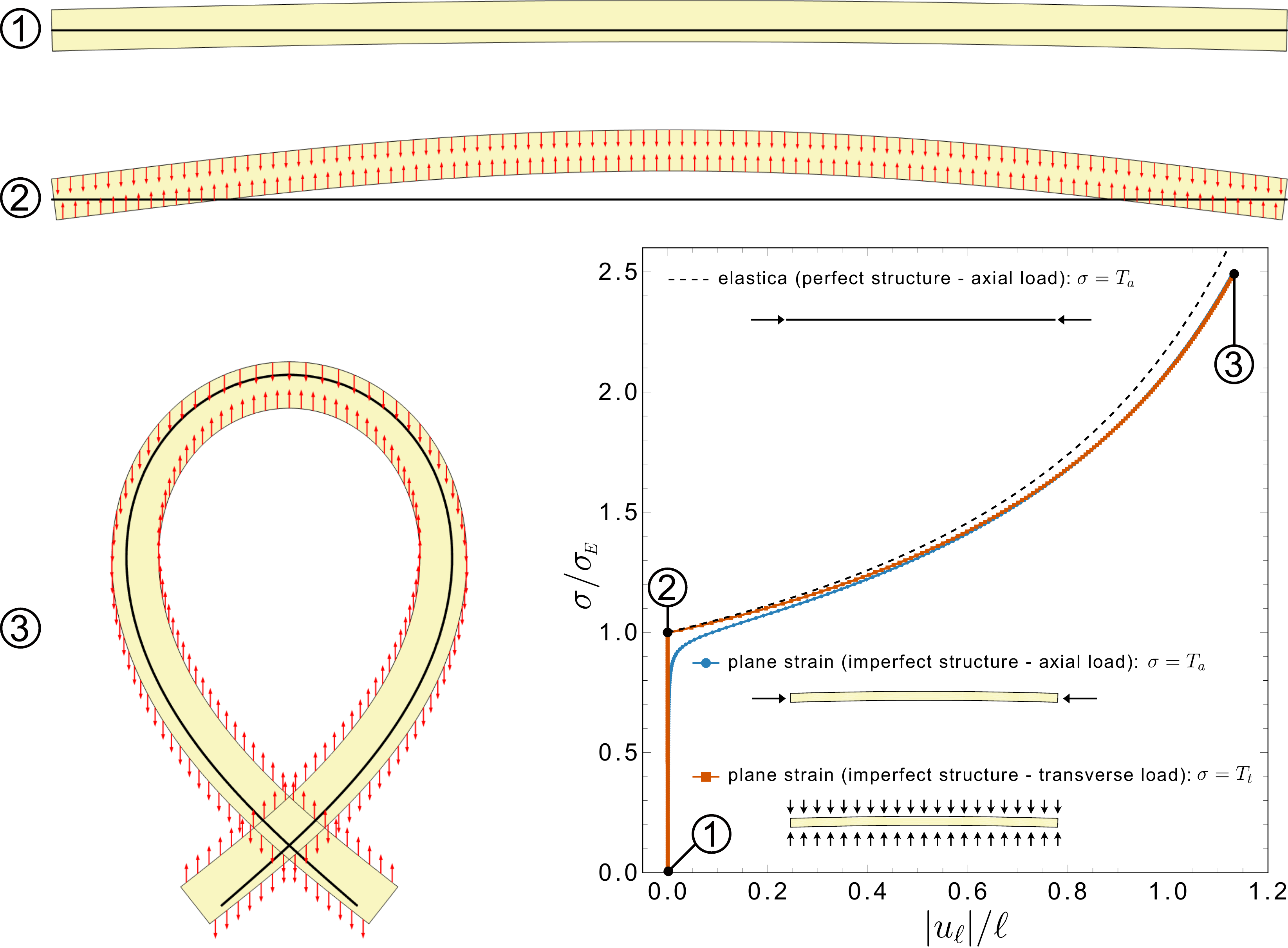}
\caption{A numerical simulation shows that a slender elastic layer subject to transverse dead forces behaves as predicted by the generalized elastica, eq.~\eqref{ocio}. 
Upper part: The elastic layer in the unloaded configuration (sketched in yellow, point 1) used for the finite element model. The initial geometric imperfection, mimicking the first buckling mode, appears as a deviation from the black line representing the perfectly straight elastica. Central part: at the critical load of the perfect elastica (still straight), the imperfect elastic layer already exhibits a finite deflection (point 2); red arrows (not to scale) denote the applied self-equilibrated transverse load. Lower left: comparison between the deformed elastic layer and the perfect elastica at $2.5$ times the critical buckling load (point 3) shows excellent agreement. Lower right: stress $\sigma$ (divided by the value at buckling) vs end displacement $u_\ell$ (divided by the initial length of the layer) for the imperfect layer (red) and the perfect elastica (black), traced up to $2.5$ times the critical load; slight deviations at higher loads arise from the different models employed (two-dimensional plane strain for the elastic layer versus one-dimensional beam formulation for the elastica); these differences diminish and eventually vanish as the slenderness of the rod increases.
}
\label{fig:postbuckling}
\end{figure}

The results of the simulations are reported in Fig.~\ref{fig:postbuckling} for the elastic layer (depicted in yellow), contrasted with the analytical solution for the perfect Euler elastica (denoted by a continuous black line).

The load-deflection curve, reported in the lower-right panel in terms of the normalized stress $\sigma/\sigma_E$, where $\sigma_E$ is the buckling load, versus the normalized end displacement $u_\ell/\ell$, confirms that the response of the perfect elastica closely matches that of the elastic layer. The small deviations observed at higher loads arise from the different modeling assumptions (plane strain versus beam theory) and disappear in the limit of infinite slenderness. The figure compares two loading cases — transverse (red curve) and axial (blue curve) — both solved using the same initial imperfection. The results show that the axial loading case is significantly more sensitive to imperfections. 

\begin{figure}[hbt!]
    \centering
    \begin{subfigure}[]{.48\textwidth}
        \centering
        \includegraphics[width=\textwidth]{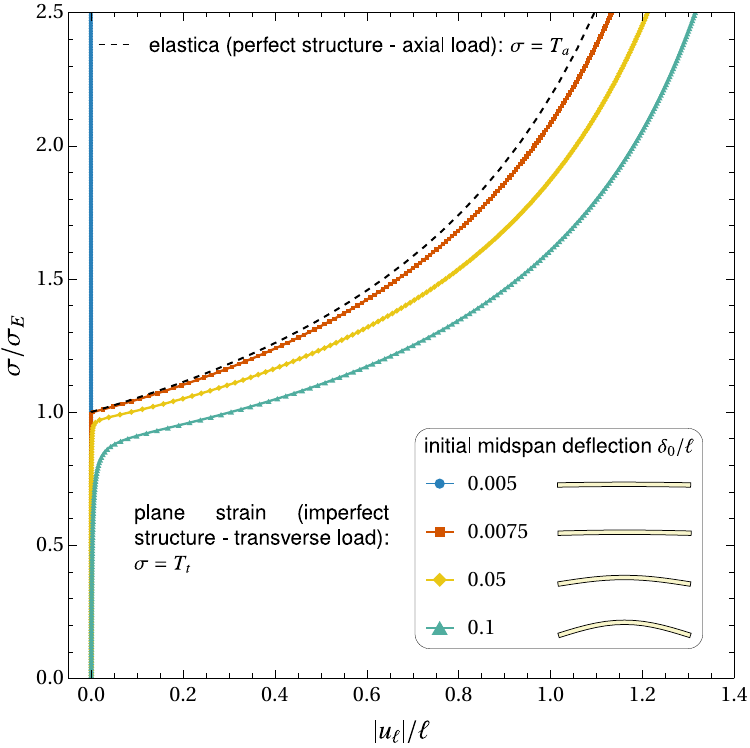}
    \end{subfigure}
    \begin{subfigure}[]{.48\textwidth}
        \centering
        \includegraphics[width=\textwidth]{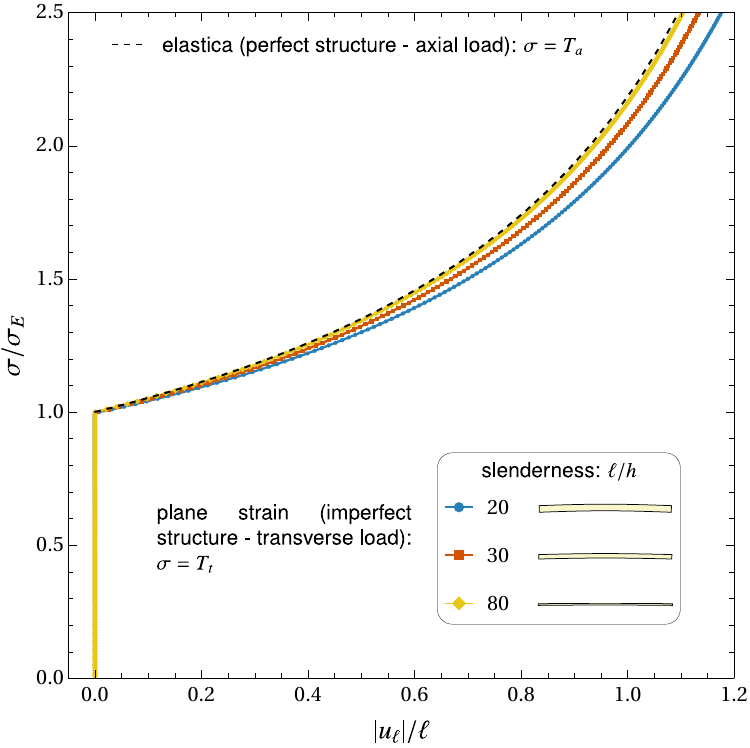}
    \end{subfigure}
    \caption{
        Finite element analyses illustrating the influence of the magnitude of initial imperfection (left) and slenderness ratio (right) on the buckling response of a rod under transverse loading. Left: load-displacement curves for different values of the normalized initial midspan imperfection $\delta_0/\ell$. For a small imperfection $\delta_0/\ell = 0.005$, transverse loading does not induce buckling and the structure follows the fundamental equilibrium path (blue curve). For larger imperfections $\delta_0/\ell \geq 0.0075$ the beam buckles, and the corresponding buckling load decreases as the imperfection increases. Right: effect of increasing slenderness on the postcritical response of a rod subject to transverse loading, showing progressive convergence of the plane-strain solution toward the Euler elastica as slenderness increases.
    }
    \label{fig:additional}
\end{figure}

Additional analyses performed with the software Comsol, reported in Fig.~\ref{fig:additional}, explore the influence of both the magnitude of the initial imperfection and the slenderness ratio. These simulations show that, for imperfections smaller than that considered in Fig.~\ref{fig:postbuckling}, the structure remains on the fundamental equilibrium path and does not buckle. In particular, an initial imperfection corresponding to a normalized initial midspan deflection $\delta_0/\ell = 0.005$ is sufficient to trigger buckling under axial loading, but not when the beam is subjected to transverse forces; in the latter case, the beam follows the fundamental equilibrium path, represented by the blue curve in the left panel of Fig.~\ref{fig:additional}. For larger imperfections, the beam loaded with transverse forces buckles, and the corresponding buckling load decreases as the imperfection increases, although this reduction is less pronounced than in the axially loaded case. Moreover, as the slenderness increases, the discrepancy between the plane strain solution and the Euler elastica, observed at higher load levels, progressively diminishes and eventually vanishes, as shown in the right panel of Fig.~\ref{fig:additional}.

The deformed configurations shown in Fig.~\ref{fig:postbuckling} provide further confirmation. In the upper panel, the unloaded configurations are superimposed, highlighting the layer's initial imperfection. The central panel corresponds to the buckling load: the perfect Euler elastica remains undeformed (straight), whereas the layer already exhibits a finite deflection due to the imperfection. The lower left panel shows the response at 2.5 times the critical load, where the configuration intersects itself and is therefore unstable. This configuration is included to demonstrate that the transversely loaded rod continues to follow the elastica equation even under very large deformations. The transverse dead forces, indicated by red arrows (not to scale), further illustrate that at extreme deformations, the load in some regions of the rod becomes oriented outward relative to the structure.

One noteworthy aspect highlighted by the numerical simulations is the low sensitivity of buckling induced by transverse loading to imperfections. In particular, the sensitivity is found to be significantly lower than that associated with buckling under axial loading.

Overall, Fig.~\ref{fig:postbuckling} provides strong validation of the analytical model, demonstrating that a transversely loaded layer exhibits, when sufficiently slender, the same response that can be predicted by the Euler elastica under axial compression.

\clearpage

\section{The design of a testing setup: experimental evidence}

The realization of an experimental setup capable of applying a transverse load, provided by dead forces, simultaneously to the extrados and intrados of an elastic rod poses a challenging problem due to several complicating factors. These are related to the fact that dead loads are defined in the reference configuration and must follow the deformation without imposing additional constraints. Moreover, the two loads applied on opposite sides of the rod must remain perfectly balanced throughout the experiment. Finally, the rod must be counterbalanced for its own weight.

All these features must be realized by designing a setup that permits the large deflections involved in the postcritical behavior of the rod. 

Therefore, it was decided to investigate eq.~\eqref{ocio2} for $n=1$ by applying a transverse {\it tensile} stress $T_t$ and measuring the corresponding buckling axial stress $T_a$. The latter measurement is carried out by comparing the experimentally observed postcritical behavior with that predicted by eq.~\eqref{ocio}, after calibrating the elastic modulus $E$ and the initial imperfection of the elastic rod (expressed in terms of the initial midspan deflection $\delta_0$) from a preliminary buckling experiment without transverse load ($T_t=0$), reported in Appendix \ref{appendicea}.

\subsection{The experimental setup for a movable transverse load}

\begin{figure}[!htb]
\centering
\includegraphics[width=\linewidth]{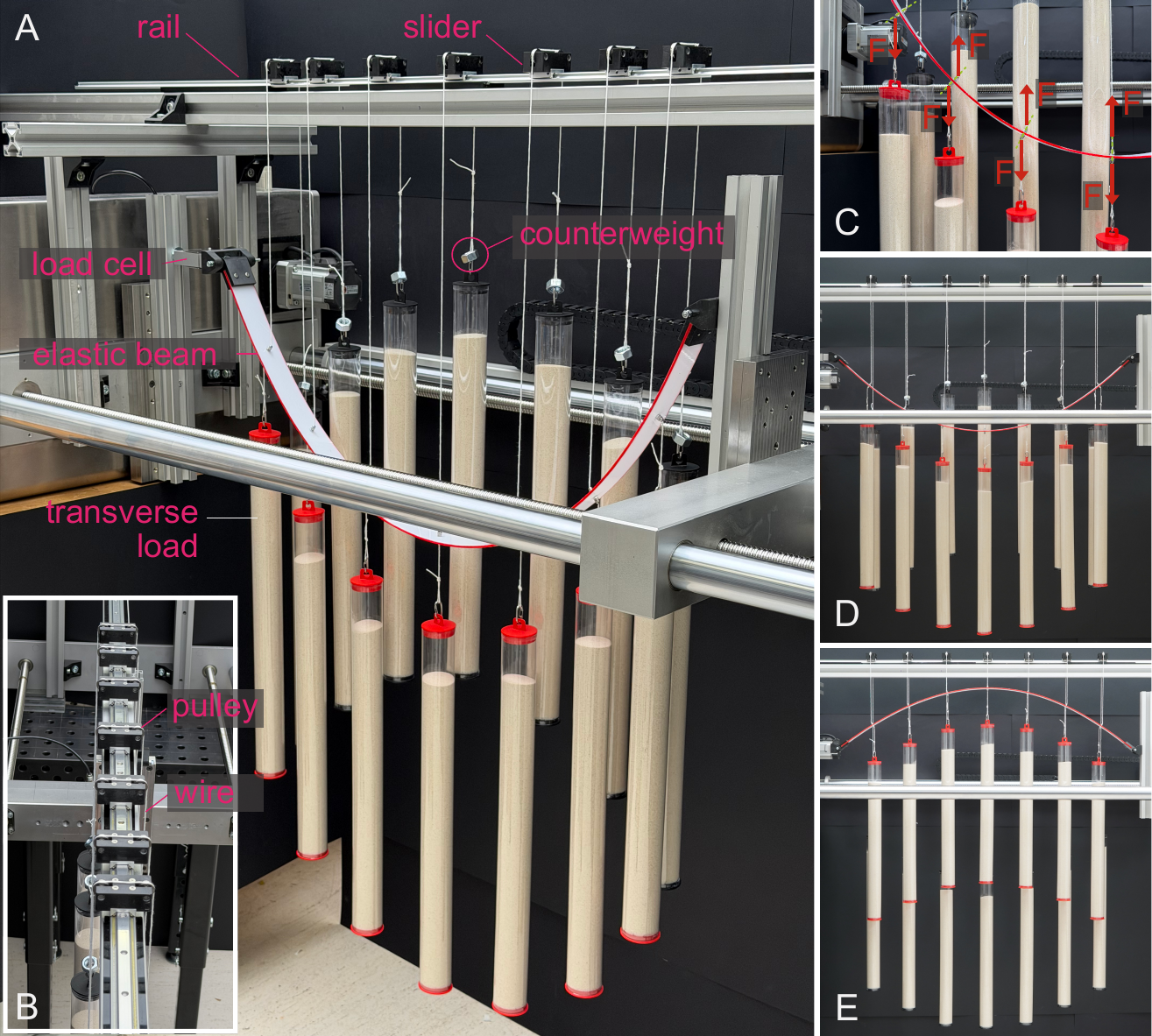}
\caption{
(A) Experimental setup during a test. (B) Detail of the rail and slider system permitting the load to move freely in the horizontal direction. The counterweight used to compensate for the rod's weight consists of a nut, while the transverse load is provided by tubes filled with sand. When the intrados loads move downward (panel D) or upward (panel E), the extrados loads move in the opposite direction—upward in panel D and downward in panel E. Panel C shows the load application system. Note the upward buckling in panel D, demonstrating the effective compensation of the rod weight.
}
\label{fig:setup}
\end{figure}

The experimental setup is shown in Fig.~\ref{fig:setup}. A polycarbonate rod was prepared with two end pins, having a total length of $\ell=1020$ mm and a rectangular cross-section of $b=40.24$ mm $\times$ $t=2.93$ mm. The rod has a radius of inertia $\rho = 0.8458$ and a slenderness $\lambda_{\text{sl}}=1205.93$. Seven retaining pins, used to suspend the cables providing the vertical load, were inserted through the thickness of the rod at equal intervals of $a=127.5$ mm. The pins increase the load application height to $h=20.21$ mm, as shown in the upper-right part of Fig.~\ref{esperi}. 

The forces applied at the intrados of the rod were generated by simply hanging weights (plastic tubes filled with sand), whereas the corresponding forces applied at the extrados were transmitted through a sophisticated pulley system specifically designed to move freely along a slider during the progressive deformation of the rod. 

The axial load $P$ was applied by imposing an increasing displacement at a rate of $1$ mm/s at the right end of the rod using an electromechanical testing machine (Messphysik Materials Testing \lq Midi 10') mounted horizontally and equipped with a DBBSM-100kg load cell manufactured by Leane International.

Fig.~\ref{fig:setup} also shows details of the sliding system that allows the vertical loads to move freely in the horizontal direction (panel B).

When the loads applied at the intrados move downward (panel D) or upward (panel E), the loads applied at the extrados move in the opposite direction, upward in panel D and downward in panel E. Panel C shows details of the rod attachment system. The self-weight of the rod was compensated by counterweights consisting of 16 g nuts applied at each loading point. The effectiveness of the weight compensation is evidenced by the occasional upward buckling of the rod, as shown in panel E, despite the axial load remaining unchanged.

\subsection{Experimental results confirming the theory}

\begin{figure}[!htb]
\centering
\includegraphics[width=\linewidth]{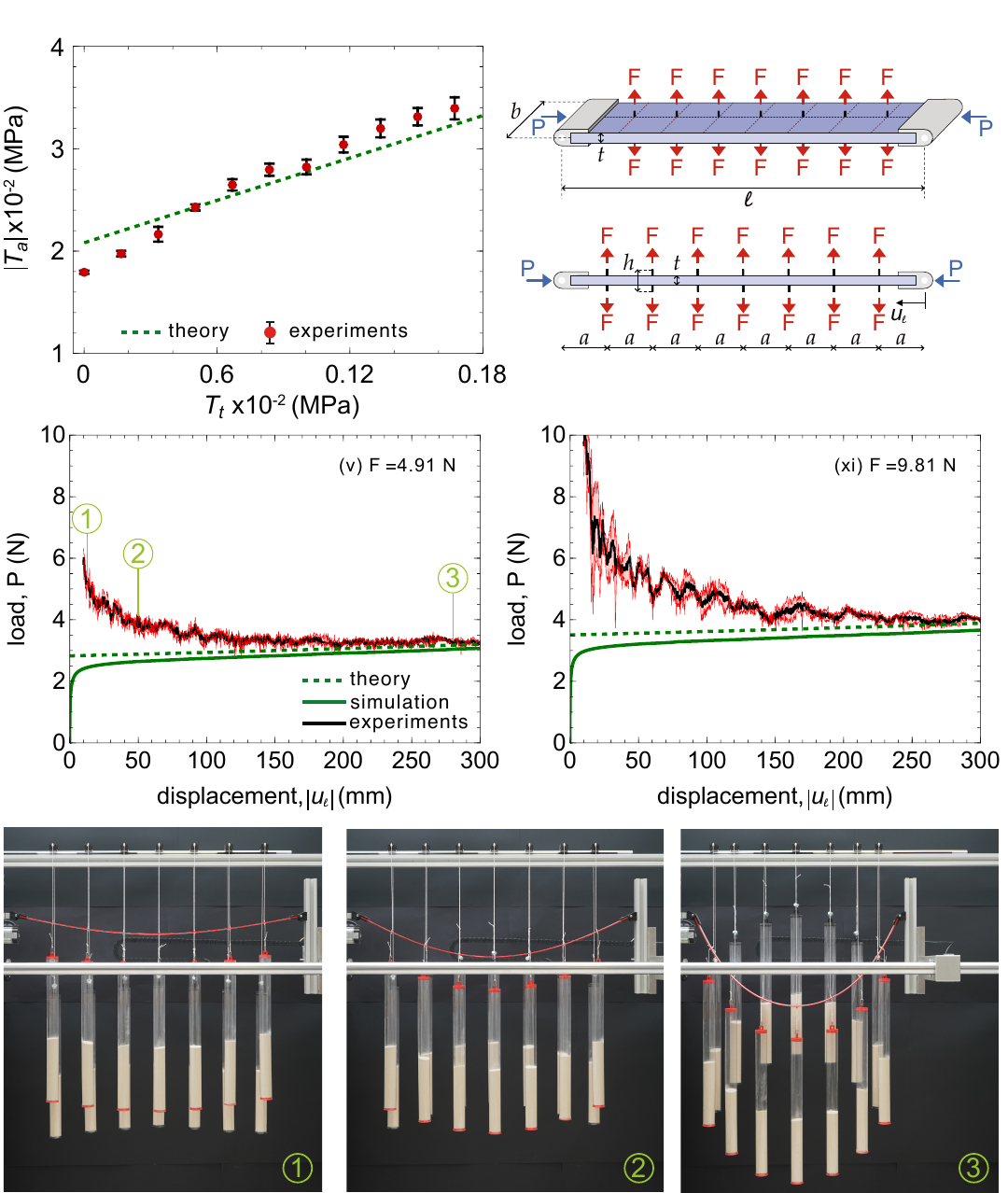}
\caption{
Upper part: experimental results (red points with error bars) demonstrating the validity of equation \eqref{ocio9}, which predicts a linear increase of the axial buckling stress $|T_a|$ with increasing transverse stress $T_t$ (a schematic of the tested rod is shown on the right). Central part: axial load $P$ versus pin displacement $u_\ell$ measured during the postcritical behavior of the rod for the two transverse loads (v) and (xi), shown as black lines. The red lines, denoting the confidence bands, are very narrow. Predictions from eq.~\eqref{ocio8} (reported dashed) and from numerical simulations (green curves) accurately describe the postcritical behavior. Lower part: three photographs of the rod at the points labeled 1-3 during postcritical deformation.
}
\label{esperi}
\end{figure}

According to eq.~\eqref{ocio2}, an increase in the transverse stress $T_t$ must correspond to a linear increase in the buckling axial stress $|T_a|$. In the experimental setup, the loads are transmitted through pins inserted through the thickness of the rod, defining a load application height $h$ that differs from the rod thickness $t$. Therefore, using eq.~\eqref{ELASTICA} with $V=0$ and $J=bt^3/12$ leads to
\begin{equation}
\label{ocio8}
    \theta'' - 12 \frac{T_a + T_t h/t}{E t^2} \sin\theta = 0, 
\end{equation}
where $T_a=P/(bt)$ and $T_t=q_2/b$, and to the corresponding buckling condition
\begin{equation}
\label{ocio9}
    T_a + T_t \frac{h}{t} = -n^2 \pi^2 \frac{E}{\lambda_{\text{sl}^2}}. 
\end{equation}
Equations~\eqref{ocio8} and \eqref{ocio9} are therefore used for comparison with the experiments. Results from the latter equation are reported with a dashed green in the upper part of Fig.~\ref{esperi} on the left, so that the intercept of this line with the vertical axis, represents the Euler buckling load, equal to $0.021$~MPa.
The experimental results, contrasted with the linear prediction obtained from eq.~\eqref{ocio9}, are reported as spots in the upper-left part of Fig.~\ref{esperi}, for 10 increments of the transverse stress, $T_t = 7F/(b \ell)$: 
\begin{center}
    \begin{tabular}{lcccccccccccc}
    \toprule
    Step $\#$ & (i) & (ii) & (iii) & (iv) &  (v) & (vi) & (vii) & (viii) & (ix) &  (x) & (xi) \\[3mm]
    $F$ [N]   &   0 & 0.98 &  1.96 & 2.94 & 3.92 & 4.91 &  5.89 &   6.87 & 7.85 & 8.83 & 9.81 \\
    \bottomrule
    \end{tabular}
\end{center}

The spots reported in the figure were obtained by averaging the load $P$ over displacements $u_\ell$ in the range 250 mm $\pm$ 20 mm. The same procedure was used to determine the confidence intervals. A truly distributed load along the rod could not be implemented experimentally; instead, seven concentrated loads $F$ were applied. The difference between these two loading conditions was assessed through Comsol simulations and was found to be negligible, so that it is not reported. 

In the central part of the same figure, two postcritical curves are shown (obtained as the mean value from three independent experiments), corresponding to transverse loads (v), 4.91~N, and (xi), 9.81~N. In the figure, the predictions obtained from eq.~\eqref{ocio8} are shown as dashed lines, while the green curves represent numerical simulations in which the load is discretized as in the experiments. 

The numerical simulations were performed using Comsol Multiphysics, following the modeling strategy described in Section~\ref{post}. In contrast to the plane-strain formulation adopted there, the experimental configuration was modeled using plane-stress elements. The finite element model includes a detailed discretization of the rod with rectangular cross-section $b \times t$, as well as the discretization of the pins of length $h$ through which the concentrated transverse loads $F$ are applied. Geometric nonlinearity is included to capture large-deflection effects, and buckling is triggered by introducing an initial geometric imperfection corresponding to the first buckling mode, scaled to match the experimentally imposed midspan deflection $\delta_0$.

The slightly higher compliance observed in the numerical simulations than in the theoretical prediction is attributed to the presence of an initial imperfection. The theoretical prediction refers to a perfect structure, whereas a small geometric imperfection was intentionally introduced in the numerical simulations, resulting in a slight reduction in structural stiffness. 

The lower part of the figure shows three snapshots of the progressive deformation of the rod under transverse load (v). The photographs correspond to horizontal pin displacement $u_\ell$ indicated by labels 1-3. Note that the transverse loads (obtained by filling cylinders with sand) move both vertically and horizontally so as to follow the deformation without introducing spurious constraints. 

Additional experimental results are reported in the Appendix \ref{appendicea}, while videos of the experiments are provided as supplementary material.

The experimental load-displacement curves in the post-critical regime of the rod are shown in black in the central part of Fig.~\ref{esperi}, with confidence bands highlighted in red.
Compared with theoretical and numerical predictions, the experimental response initially appears stiffer and gradually approaches the expected behavior as the displacement increases. A displacement of 150 mm is sufficient to achieve good agreement; for this reason, and to minimize frictional effects, the data reported in the upper-left part of Fig.~\ref{esperi} were obtained by averaging the load $P$ over displacements recorded within the range 250 mm $\pm$ 20 mm.
The initial discrepancy is attributed to the fact that, despite all possible measures taken to minimize friction, the system must overcome an initial detachment friction arising from the multiple elements involved in the rod deformation. Taking this into account, the experiments confirm the theoretical predictions and demonstrate that transverse dead loads of opposite sign can be effectively realized in practice.

\section{Conclusion}
\label{conclu}

A doubly supported elastic rod has been investigated, straight in its initial configuration and loaded with two equal and opposite transverse loads applied orthogonally to its axis and uniformly distributed along its length. Although such a dead load distribution is generally assumed to leave the rod unaffected and has consequently not been considered, it is shown that the structure behaves as if it were axially loaded and consequently obeys a generalized form of Euler elastica, where the axial stress adds to the transverse stress. 
Multiple independent theoretical and numerical analyses have been provided in support of this result. Crucially, dedicated experiments were also designed and performed, not only validating the theoretical predictions but also demonstrating that the proposed transverse loading can be realized in practice.
These findings shed new light on the mechanical effects of transverse loading on structural elements and may find applications in the mechanics of slender bodies, beams, and filaments.

\section*{Acknowledgements}
D.B. and A.P. acknowledge funding from the European Research Council (ERC) under the European Union’s Horizon Europe research and innovation programme, Grant agreement No. ERC-ADG-2021-101052956-BEYOND. 
D.M acknowledges financial support from the European Union, ERC grant HE GA 101086644 S-FOAM. 
The methodologies developed in this work fall within the aims of the GNFM (Gruppo Nazionale per la Fisica Matematica) of the INDAM (Istituto Nazionale di Alta Matematica). 
Views and opinions expressed are those of the authors only and do not necessarily reflect those of the European Union or the European Research Council Executive Agency. Neither the European Union nor the granting authority can be held responsible for them.

\printbibliography

\appendix
\section{Appendix: Further details on experiments}
\label{appendicea}

\begin{figure}[!htb]
\centering
\includegraphics[width=\linewidth]{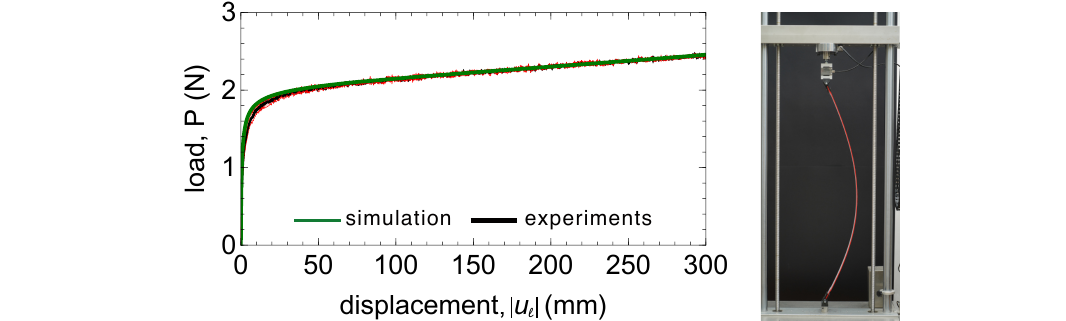}
\caption{
Initial buckling tests of the rod used in all subsequent experiments. Left: estimation of the elastic modulus $E=2685$ MPa obtained by matching the results of three buckling experiments (mean value shown as a black line, confidence bands in red) with a Comsol numerical simulation. The curves are superimposed. Right: photograph taken during a test.
}
\label{fig:euler:taratura}
\end{figure}

The calibration of the rod's elastic modulus $E$ for subsequent comparison with experiments was carried out by matching a numerical simulation of the rod's bifurcation under compression (performed in Comsol) to three experiments conducted on the polycarbonate bar used in the subsequent tests. The comparison is shown in Fig.~\ref{fig:euler:taratura}, where the experimental confidence bars are indicated in red. The curves were found to be practically superimposed with the estimated Young's modulus 
$E=2685$ MPa, which leads to an Euler critical load equal to $2.148$ N. 
The initial geometric imperfection was determined from the same set of experiments. It corresponds to the first buckling mode, with an initial midspan deflection of $\delta_0 = 0.0102$ m. 

All experiments, including the initial calibration test, were repeated three times, and the results are reported as mean values. No additional repetitions were required due to the excellent repeatability of the measurements, as evidenced by the narrow error bars.

\begin{figure}[!htb]
\centering
\includegraphics[width=\linewidth]{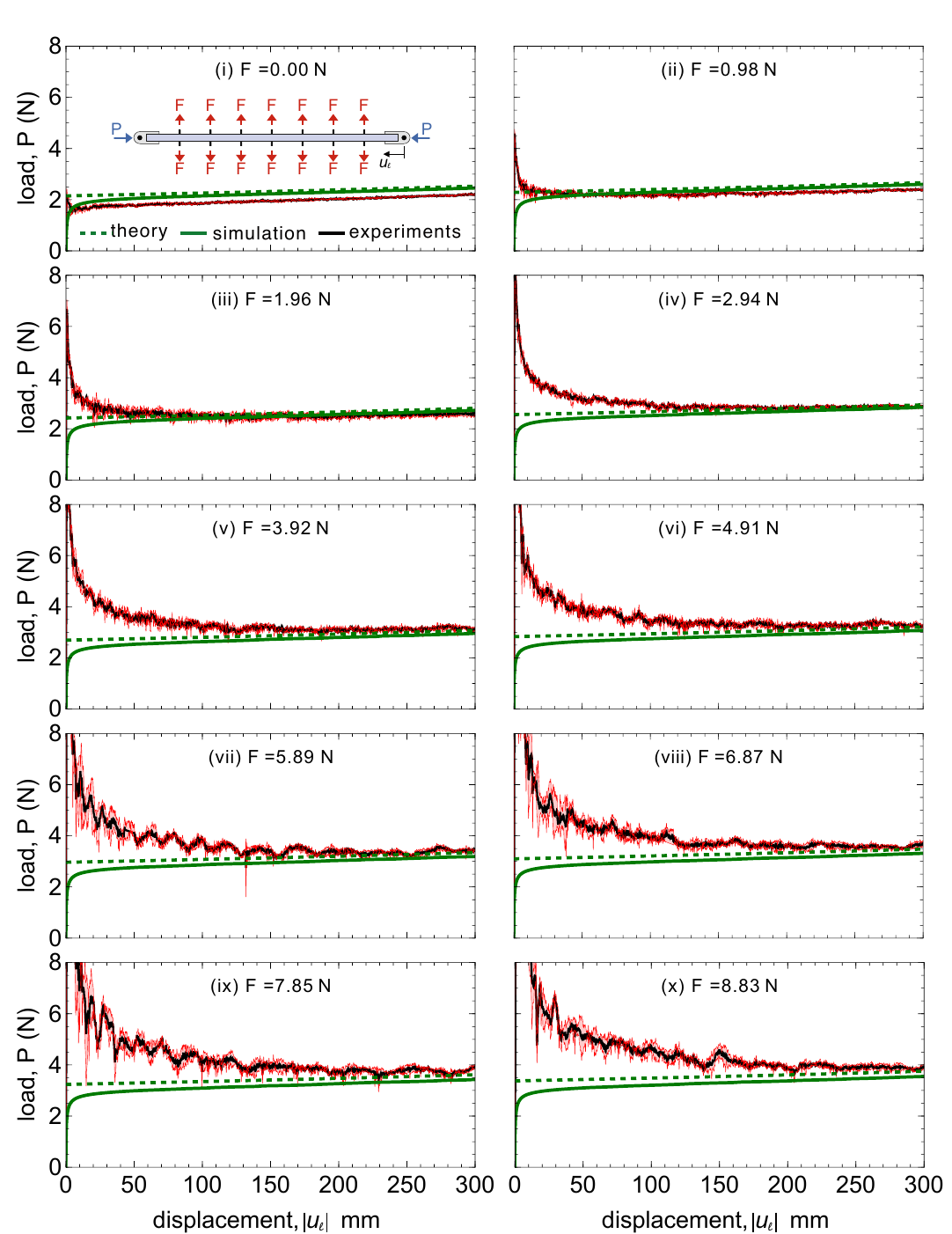}
\caption{
Confirmation of the theoretical predictions from eq.~\eqref{ocio8} through axial load $P$ versus pin displacement $u_\ell$ curves (black line), measured during the postcritical behavior of the rod for transverse loading steps (i)--(x). The red lines denote the confidence bands. Predictions from eq.~\eqref{ocio8} (reported dashed) and from numerical simulations (green curves) accurately describe the postcritical behavior.
}
\label{fig:results:complessivo}
\end{figure}

Results analogous to those shown in the central part of Fig.~\ref{esperi} are reported in Fig.~\ref{fig:results:complessivo}, which collects all loading steps (i)--(x), with the exception of step (xi), already included in Fig.~\ref{esperi}. In the figure, the mean experimental response is shown as a black curve, while the red lines denote the confidence bands. The latter are very narrow, further confirming the high quality and repeatability of the experimental data. 

\begin{figure}[!htb]
\centering
\includegraphics[width=0.90\linewidth]{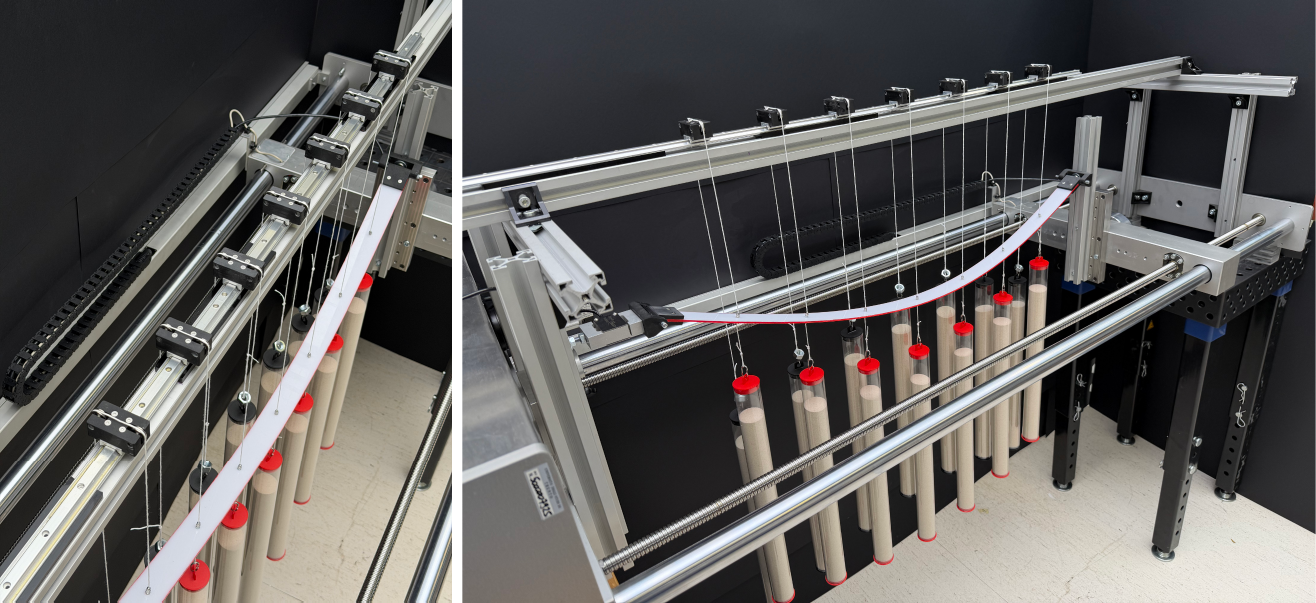}
\caption{
Two lateral views of the experimental setup during the transverse loading step (v), showing the pulley-slider system used to allow the loads to move freely. The electromechanical loading machine, rotated to a horizontal position, is also visible.
}
\label{fig:setup:details}
\end{figure}

Allowing the transverse load to follow the deformation represents one of the most challenging aspects of the experimental setup. Figure~\ref{fig:setup:details} shows two lateral views of the setup during operation, highlighting the system of pulleys and sliding guides employed to ensure the desired mobility of the vertical loads.

\begin{figure}[!htb]
\centering
\includegraphics[width=0.90\linewidth]{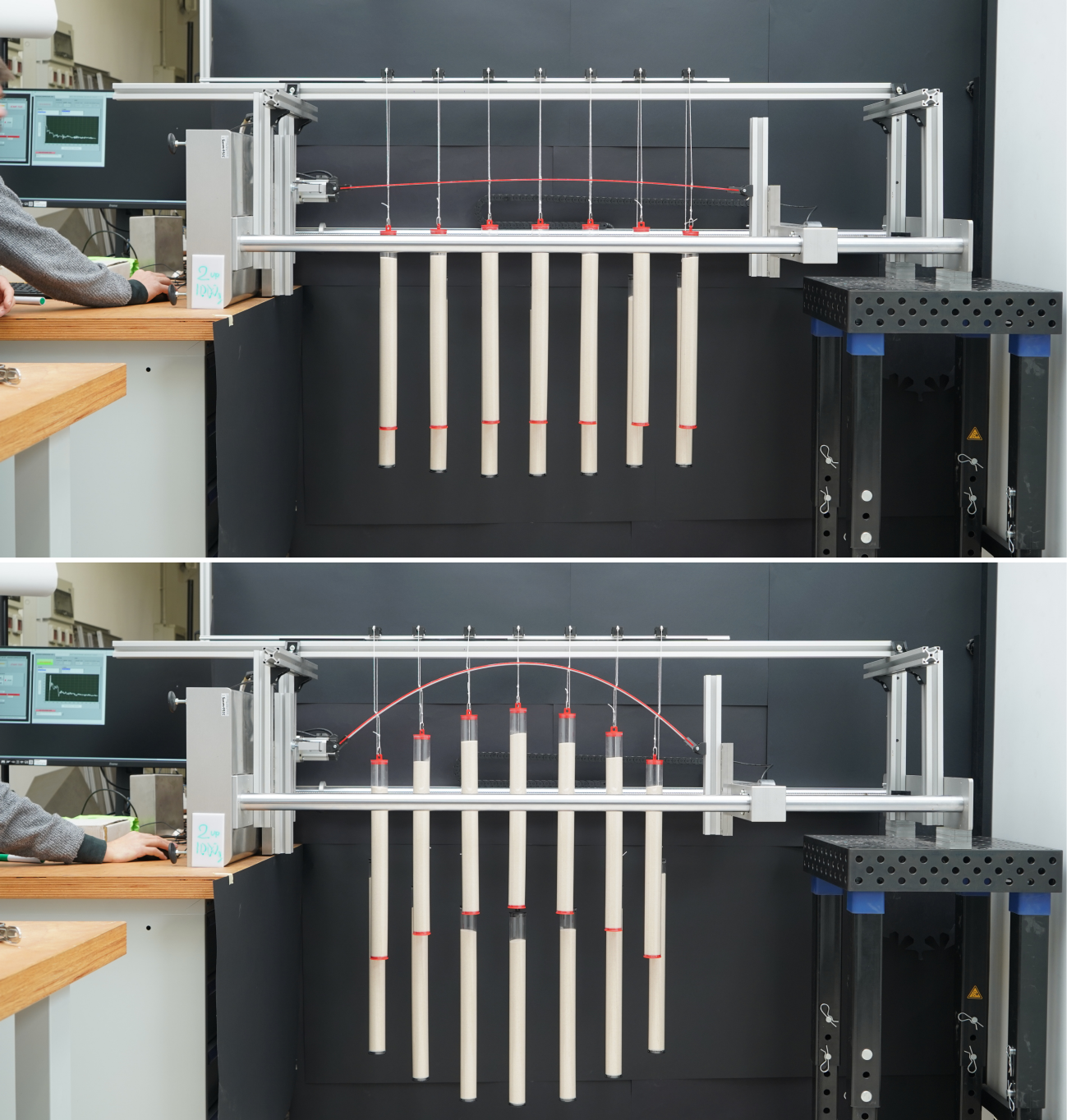}
\caption{
Two photographs showing the initial (upper part) and late (lower part) stages of an upward buckling event, indicating effective compensation of the rod's self-weight. 
}
\label{fig:setup:frontal:view}
\end{figure}

The effective compensation of the rod’s self-weight is demonstrated by the occasional occurrence of upward buckling while the axial load remains unchanged. This behavior is documented in Fig.~\ref{fig:setup:frontal:view}, corresponding to the transverse loading step (xi). The upper image shows the initial configuration, while the lower image captures a later stage of deformation.

Additional experimental material is provided in the Supporting Information in the form of videos.

\end{document}